\documentclass{pasj00}
\draft

\begin{document}
\SetRunningHead{Kimura S. \& Tsuribe T.}{The Condition of GI in Protoplanetary Disk}
%\Received{}%{yyyy/mm/dd}
%\Accepted{}%{yyyy/mm/dd}
%\Published{}%{yyyy/mm/dd}

\title{Conditions for Gravitational Instability in Protoplanetary Disks}

%%% begin:list of authors
% Do NOT capitalize all letters in "textsc".
\author{Shigeo S. \textsc{Kimura} and Toru \textsc{Tsuribe}} %
%%  \thanks{Example: Present Address is xxxxxxxxxx}}
\affil{Department of Earth and Space Science, Graduate School of Science, Osaka University,\\
 1-1 Machikaneyama-cho, Toyonaka-shi, Osaka 560-0043}
\email{kimura@vega.ess.sci.osaka-u.ac.jp , tsuribe@vega.ess.sci.osaka-u.ac.jp }

%%\and
%%
%%\author{Toru \textsc{Tsuribe}}
%%\affil{B-Address of Institute}\email{bbbbb@xxx.xxx.xx.xx}

%%\author{C-Firstname {\sc C-Familyname}}
%%\affil{C-Address of Institute}\email{ccccc@xxx.xxx.xx.xx}
%%% end:list of authors

%%% Please use the following style in case that sorting by 
%%% affiliation is impossible. 
%
% \author{%
%   D-Firstname \textsc{D-Familyname}\altaffilmark{1}
%   E-Firstname \textsc{E-Familyname}\altaffilmark{1,2}
%   and
%   F-Firstname \textsc{F-Familyname}\altaffilmark{2}}
% \altaffiltext{1}{Address of Institute}
% \email{ddddd@xxx.xxx.xx.xx}
% \email{eeeee@xxx.xxx.xx.xx}
% \altaffiltext{2}{Address of Institute}

%% `\KeyWords{}' always has to be placed before `\maketitle'.
 %Do NOT move this preamble from here!

\maketitle

\begin{abstract}
Gravitational instability is one of considerable mechanisms to explain the formation of giant planets. 
We study the gravitational stability for the protoplanetary disks around a protostar. 
The temperature and Toomre's Q-value are calculated by assuming local equilibrium between viscous heating and radiative cooling (local thermal equilibrium). 
We assume constant $\alpha$ viscosity and use a cooling function with realistic opacity. 
Then, we derive the critical surface density $\Sigma_{\rm{c}}$ that is necessary for a disk to become gravitationally unstable as a function of $r$. 
This critical surface density $\Sigma_{\rm c}$ is strongly affected by the temperature dependence of the opacity. 
At the radius $r_{\rm c}\sim 20$AU, where ices form, the value of $\Sigma_{\rm c}$ changes discontinuously by one order of magnitude. 
This $\Sigma_{\rm c}$ is determined only by local thermal process and criterion of gravitational instability. 
By comparing a given surface density profile to $\Sigma_{\rm c}$, one can discuss the gravitational instability of protoplanetary disks. 
As an example, we discuss the gravitational instability of two semi-analytic models for protoplanetary disks. 
One is the steady state accretion disk, which is realized after the viscous evolution. 
The other is the disk that has the same angular momentum distribution with its parent cloud core, which corresponds to the disk that has just formed. 
As a result, it is found that the disks tend to become gravitationally unstable for $r\ge r_{\rm c}$ because ices enable the disks to become low temperature. 
In the region closer to the protostar than $r_{\rm c}$, it is difficult for a typical protoplanetary disk to fragment because of the high temperature and the large Coriolis force. 
From this result, we conclude that the fragmentation near the central star is possible but difficult. 
%%Since $\Sigma_{\rm c}$ is determined by the local thermal equilibrium and the unstable condition, $\Sigma_{\rm c}$ is available to evaluate the possibility of disk fragmentation, regardless of the process of disk formation. 
%%Since $r_{\rm c }$ is a few tens AU, we suggest that it is unlikely for a protoplanetary disk to fragment at $r<10$AU from the central star. 
%%we suggest that the fragmentation of the disk near the central star is difficult.
\end{abstract}

\section{Introduction}
Two major models are currently considered for the formation of gas giant planets. 
One is the core accretion model (CA model; \cite{Gol73}; \cite{Hay81}; \cite{Pol96}), and the other is the gravitational instability model (GI model; \cite{Cam78}; \cite{Dur07}). 
In CA model, massive solid cores are made from dust first, and then gas giant planets are formed by the gas accretion onto these cores. 
In GI model, a disk around a protostar fragments into pieces by gravitational instability and these pieces become gas giant planets. 
Recently, extrasolar planets are discovered by direct imaging (\cite{Mar08}; \cite{Kal08}). 
These planets are heavier than Jupiter and farther from the central star than Neptune. 
Unless some additional ideas are considered, it is difficult for CA model to explain these planets because it is thought that the gas disappears from the disk before the formation of the massive solid core \citep{Dod09}. 
On the other hand, GI model has the possibility to make gas giant planets far from the central star if the mass of a disk is greater than that of the protostar. 
Thus, GI model attracts attention to explain their formation. 

In order to understand planet formation by GI model, it is desirable to know the physical condition for disk fragmentation and the position where the fragmentation occurs. 
At present, two criteria are frequently used to discuss whether a protoplanetary disk is likely to fragment. 
The first is Toomre's stability criterion \citep{Too64}
\begin{equation} 
Q \equiv \frac{c_{\rm_s}\kappa_{\rm{ep}}}{\pi G\Sigma} > 1, \label{ToomreQ} \\ 
\end{equation} 
with gravitational constant $G$, epicyclic frequency $\kappa_{\rm{ep}}$, sound speed $c_{\rm{s}}$, surface density $\Sigma$, and stability parameter $Q$. 
\citet{Too64} showed that the infinitesimally thin disk is stable if the stability criterion (\ref{ToomreQ}) is satisfied. 
It is necessary to violate this criterion in order for fragmentation to occur. 
The other criterion is Gammie's cooling criterion \citep{Gam01} 
\begin{equation} 
\beta \equiv t_{\rm{cool}}\Omega=u\left(\frac{du}{dt}\right)^{-1}\Omega\lesssim\beta_{\rm c}, \label{Gammie'sBeta} 
\end{equation} 
where $\Omega$ is the angular velocity, $t_{\rm{cool}}=u\left(du/dt\right)^{-1}$ is the cooling time, $u$ is the specific internal energy, $du/dt$ is the total cooling rate, $\beta$ is cooling parameter, and $\beta_{\rm c}$ is the critical cooling parameter. 
\citet{Gam01} suggested that rapid cooling is necessary for fragmentation, in addition to violating the Toomre criterion (\ref{ToomreQ}). 
Using shearing sheet simulations with constant cooling rate and without heating except for artificial viscosity, he showed that a self-gravitating disk can fragment only if $\beta\lesssim 3$. 
If not, disk cannot fragment because the disk can be stabilized by internal heating due to the turbulence arising from gravitational instability. 
This state is called "gravitoturbulence". 
It is said that the heating owing to the gravitoturbulent dissipation automatically controls the Q value close to unity. 
The gravitoturbulence has the potential to stabilize the system where Toomre criterion is violated, while it does not arise in the disk that satisfies Toomre criterion. 

The $\beta_{\rm c}$ obtained by \citet{Gam01} is for the specific case (local, 2D, etc.). 
Many three-dimensional simulations were carried out in order to obtain the general value of $\beta_{\rm c}$ (e.g., \cite{Ric05}; \cite{Cla07}; \cite{Cos10}). 
However, each simulation suggested the different $\beta_{\rm c}$, and general $\beta_{\rm c}$ has not been clarified yet. 
Furthermore, it is not obvious whether or not $\beta_{\rm c}$ exists as some value. 
\citet{Mer11a} calculated the disks with different initial parameters such as the disk size, the surface density profile, the mass of the disk, and the mass of the central star. 
From their calculations, it is found that $ \beta_{\rm c}  $ is different among the disks with a different set of initial parameters. 
This result shows that it is not determined only by $\beta$ whether the disk can fragment or not. 
Moreover, \citet{Mer11b} calculated the $\beta_{\rm c}$ with various numerical resolutions. 
They showed that the disk was able to fragment with larger $\beta$ if the resolution is better, and they had no indication of the convergence of $\beta$. 
From these results, it is not clear that Gammie criterion is always applicable to discuss the fragmentation of a disk. 

Some analytical works for disk fragmentation have been done. 
\citet{Raf05} argued that a protoplanetary disk is unlikely to fragment near the protostar ($r\lesssim100$AU) by using two criteria described above. 
However, as noted above, Gammie criterion is uncertain, and thus this argument includes the same uncertainty. 
On the other hand, Toomre criterion is probably assured in the sense that the disk violating Toomre criterion is gravitationally unstable. 
Other previous analytical works for disk fragmentation (e.g., \cite{Cla09}; \cite{Kra11}) assumed gravitoturbulent disks with $Q=1$. 
These studies do not have the ability to discuss the position where Toomre criterion is satisfied. 
Since Toomre criterion provides robust necessary condition for fragmentation, the studies based solely on Toomre criterion (\ref{ToomreQ}) is favorable to know the possibility of fragmentation. 

In this study, we investigate the gravitational stability in protoplanetary disks based solely on Toomre criterion without the uncertainty associated with Gammie criterion. 
We do not aim to consider the time evolution of particular disks but derive an instability condition that is available for various disks. 
Since recent observation shows the diversity of protoplanetary disks and planetary systems, we think that it is important to derive such a widely useful formulation. 
In order to calculate Toomre's Q value with given $\Sigma$ and $\kappa_{\rm{ep}}$, it is necessary to know the temperature. 
Thus, it is significant to consider realistic thermal processes. 
We construct an analytical model for temperature calculation with thermal process in section \ref{sec:TEMP}. 
In section \ref{sec:CSD}, we derive the critical surface density $\Sigma_{\rm c}$ that is necessary for a disk to become gravitationally unstable. 
This $\Sigma_{\rm c}$ is available to discuss the possibility and position of gravitational instability, regardless of surface density profile or formation process of the disks. 
In section \ref{sec:APP}, we introduce two semi-analytic models for protoplanetary disks, and discuss the possibility and location of the gravitational instability by comparing the surface densities of these models to $\Sigma_{\rm c}$. 
In section \ref{sec:DIS}, our results are discussed and compared to previous studies. 
Finally, we summarize this study in section \ref{sec:CON}.

\section{Temperature}\label{sec:TEMP}
In this section, an analytic model for the temperature calculation is introduced. 
The local thermal equilibrium between viscous heating and radiative cooling are assumed. 
We use Keplerian rotation law, ideal gas, vertical hydrostatic equilibrium, and a geometrically-thin disk. 
We assume that radiation escapes only in the vertical direction.
Effective viscosity with constant $\alpha$ introduced by \citet{Sha73} is adopted. 
Since these assumptions are similar to the standard accretion disk \citep{Pri81}, we have 
\begin{eqnarray}
\Omega&=&\sqrt{GM_{\rm s}/r^3} ,\label{Omega}\\
c_{\rm s}^2&=&\gamma k_{\rm B}T/\overline m , \label {cs}\\
H&=&c_{\rm_s}/\Omega, \label{H}\\
\rho&=&\Sigma/2H=\Sigma\Omega/2c_{\rm s},\\
\Gamma&=&\frac{9}{8}\alpha c_{\rm s}H\Sigma\Omega^2 ,\label {Q+}\\
\Lambda &=& \left\{
\begin{array}{ll}
{\displaystyle \frac{32\sigma T^4}{3\kappa\Sigma}}\ & (\tau \ge 1)  \\[10pt]
{\displaystyle \frac{8\sigma T^4\kappa\Sigma}{3}}\ &  (\tau <1)\ \\
\end{array}
\right. \ ,\label{Q-}\\
\mbox{and }\nonumber \\
\Gamma&=&\Lambda, \label{LTE}
\end{eqnarray}
where $\gamma$ is specific heat, $\overline m$ is mean molecular weight, $\rho$ is density, $H$ is scale height, $M_{\rm s}$ is the mass of protostar, $\sigma$ is Stefan-Boltzmann constant, $k_{\rm B}$ is the Boltzmann constant, $\Gamma$ is viscous heating rate, $\kappa$ is opacity, $\Lambda$ is cooling rate per unit area, and $\tau=\Sigma\kappa/2$ is optical depth of vertical direction. 
We approximate cooling rate by connecting optically thin and thick limit continuously. 
Here, we use the Rosseland mean opacity $\kappa(\rho,T)$ that is approximated by using power-law as
\begin{equation}
\kappa=\kappa_0\rho^aT^b ,\label{BellLin}
\end{equation}
where the values of $\kappa_0,\ a,\ b$ are summarized in table \ref{tab:BellLin} (\cite{Bel94}; \cite{Cos10}). 
This table shows the characteristic temperatures at which the main component of opacity changes. 
For example, $T_1=166.8\rm K$ is the temperature where the main component of opacity changes between ices and sublimation of ices, and $T_2=202.6\rm K$ is between sublimation of ices and dust grains.
Other characteristic temperatures depend on the density.

\begin{longtable}{llllll}
	\caption{Bell\&Lin opacity} \label{tab:BellLin}
  \hline              
 &  &  &  & Temperature & Temperature \\ 
Opacity regime & $\kappa_0(\rm cm^2/g)$ & $a$ & $b$ & range From(K) & range To(K)
\endfirsthead
  \hline
\endlastfoot
  \hline
  \hline
Ices & $2\times10^{-4}$ & 0 & 2 & 0 & 166.8 \\
Sublimation of ices & $2\times10^{16}$ & 0 & -7 & 166.8 & 202.6 \\
Metal dust & $1\times10^{-1}$ & 0 & 1/2 & 202.6 & 2286.7$\rho^{2/49}$ \\
Sublimation of metal dust & $2\times10^{81}$ & 1 & -24 & 2286.7$\rho^{2/49}$ & 2029.7$\rho^{1/81}$ \\
Molecules & $1\times10^{-8}$ & 2/3 & 3 & 2029.7$\rho^{1/81}$ & 10000.0$\rho^{1/21}$\\
Hydrogen scattering & $1\times10^{-36}$ & 1/3 & 10 & 10000.0$\rho^{1/21}$ & 31195.2$\rho^{4/75}$ \\ \hline
\end{longtable}

In order to calculate the temperature as a function of $r$ and $\Sigma$, seven variables $\Omega,\ c_{\rm s},\ H,\ \rho,\ \kappa,\ \Gamma,\ \Lambda$ can be eliminated by using the eight equations from (\ref{Omega}) to (\ref{BellLin}). 
Thus, we can obtain the equilibrium temperature analytically for a given $r$ and $\Sigma$. 
Additionally, we introduce the minimum temperature $T_{\rm{min}}$ in order to mimic the effect of ambient heating sources in a molecular cloud. 
If the above equilibrium temperature becomes lower than $T_{\rm{min}}$, we simply use $T_{\rm{min}}$ instead of the equilibrium temperature. 
As a result, the temperature can be represented as 
\begin{eqnarray}
T=\left\{
\begin{array}{ll}
{\displaystyle \left(\frac{27}{256}\kappa_0AB^ar^{-3(a+1)/2}\Sigma^{a+2}\right)^{1/(3-b+a/2)}} & (\tau \ge 1)  \\[10pt]
{\displaystyle \left(\frac{27A}{64\kappa_0B^a}r^{3(a-1)/2}\Sigma^{-a}\right)^{1/(3+b-a/2)}} & (\tau <1)  \\[10pt]
T_{\rm{min}}  & (T\le T_{\rm{min}})\ 
\end{array}
\right., \label{temp}
\end{eqnarray}
where
\begin{eqnarray}
A&=&\frac{\alpha\gamma k_{\rm B}\sqrt{GM_{\rm s}}}{\overline m \sigma} \ \label{A} \\
\mbox{and }\nonumber \\
B&=&\frac{1}{2}\sqrt{\frac{\overline m GM_{\rm s}}{\gamma k_{\rm B}}}\ .\label{B}  \nonumber
\end{eqnarray}
From equation (\ref{temp}), we can calculate temperature as a function of $\Sigma$ and $r$ for a given $\alpha$, $M_{\rm s}$, and $T_{\rm{min}}$. 
In this study, we set $\alpha=0.01$, which is the typical value of $\alpha$ caused by MRI (magnet rotational instability) turbulence, and assume $T_{\rm{min}}=10\rm K$, which is the typical value of a molecular cloud. 

\begin{figure}[tbp]
	\begin{center}
\FigureFile(80mm,50mm){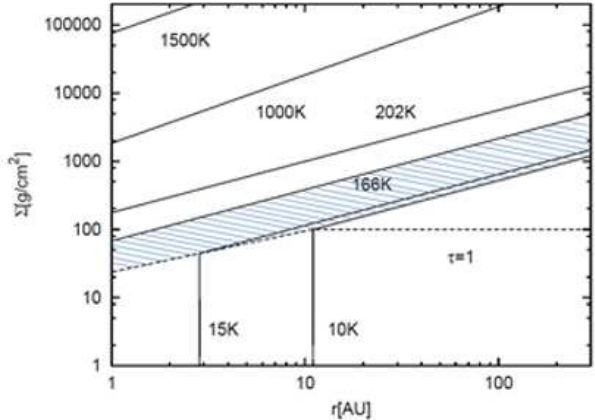}	
	\caption{Contours of the temperature for the case with $M_{\rm s}=1\rm M_{\odot}$ and $\alpha=0.01$ and $T_{\rm{min}}=10\rm K$. The solid lines show contours of temperature. The dotted line represents $\tau=1$.}
		\label{fig:Tcntr}
	\end{center}
\end{figure}
 
Figure \ref{fig:Tcntr} shows the temperature in the $r-\Sigma$ plane for the case with $M_{\rm s}=1\rm M_{\odot}$, $\alpha=0.01$ and $T_{\rm{min}}=10\rm K$ as the fiducial case in this study. 
The solid lines show the contours of temperature, and the dotted line shows the $\tau=1$ line. 
The characteristic temperature $T_1$ defined above is labeled as 166K, and $T_2$ is labeled as 202K. 
The line for 166K can be regarded as snow line because ices become the main component of the opacity in the area below this line. 
Since other characteristic temperatures depend on the density, they do not coincide with contour lines of iso-temperature. 
At these characteristic temperatures, the property of opacity drastically changes as shown in equation (\ref{BellLin}) and table \ref{tab:BellLin}. 
This property of opacity has much influence on the gravitational stability of protoplanetary disks. 

From figure \ref{fig:Tcntr}, it is seen that the temperature is high for large $\Sigma$ and small $r$. 
Above dotted line, the temperature is high for large surface density because a large amount of matter prevents radiation from escaping the system. 
In a small radius, the dynamical time scale is short, and thus the heating rate is large. 
This is why the temperature is high for large $\Sigma$ and small $r$. 
The property of cooling rate drastically changes whether or not the condition $\tau\ge1$ is satisfied as described in equation (\ref{Q-}). 
Below the dotted line, which means that the disk is optically thin, it is seen that the temperature is independent of surface density. 
This is because all radiation emitted from disk can escape in the optically thin case. 
At the right-bottom area of the line with 10K, the temperature becomes the minimum temperature $T_{\rm{min}}$ due to the small viscous heating rate and large cooling rate. 
We henceforth name this area ``isothermal region". 
Note that the dotted line exists only at low temperature area, such that $10\rm K\le T\le20\rm K$. 
This is due to the high cooling efficiency for the disk with $\tau\sim 1$.

\if0
The dotted line slopes upward when going left to right until the line enters isothermal zone. 
Using the condition $\tau=\Sigma\kappa/2=1$ and ices opacity, we gain
\begin{equation}
\Sigma_{\tau}=\left\{\left(\frac{128}{27}\right)^2 10^{11} \frac{r^3}{A^2}\right\}^{1/5}=24\ {\rm g/cm}^2\ \left(\frac{\alpha}{0.01}\right)^{-2/5}\left(\frac{M_{\rm s}}{1\rm M_{\odot}}\right)^{-1/5}\left(\frac{r}{1\rm{AU}}\right)^{3/5}\ .
\end{equation}
Ices opacity is increasing function of the temperature in 10K$<T<$166K, and temperature is small with large $r$. 
Thus, large $\Sigma$ is needed for satisfying $\tau=1$ at large radius. 
\fi

%%------------------------------------------------------------------------------------------------

\section{Critical Surface Density}\label{sec:CSD}
Now we can calculate Toomre's $Q(r,\Sigma)$ by using equations (\ref{ToomreQ}) and (\ref{temp}). 
The analytical expression for Q value is summarized in appendix \ref{app:TCSD}. 
In figure \ref{fig:Qcntr}, the solid line represents the critical surface density $\Sigma_{\rm c}$ that is necessary to satisfy $Q=1$ for a given $r$. 
As references, the dotted and dashed lines represent the surface densities that satisfy $Q=0.5$ and $Q=2$, respectively. 
A disk is expected to become gravitationally unstable if the disk has a larger surface density than $\Sigma_{\rm c}$. 
This critical surface density $\Sigma_{\rm c}$ is determined only by local thermal process and criterion for gravitational instability. 
Thus, one can discuss the possibility and position of gravitational instability regardless of disk formation process by comparing a given surface density profile to $\Sigma_{\rm c}$. 
We show examples of this discussion in the next section. 
However, one should keep in mind some cautions, for example, 1) gravitational instability is not the same as fragmentation and 2) Q value is not universal for the case with non-uniform medium or with non-axisymmetric situations. 

\begin{figure}[btp]
	\begin{center}
\FigureFile(80mm,50mm){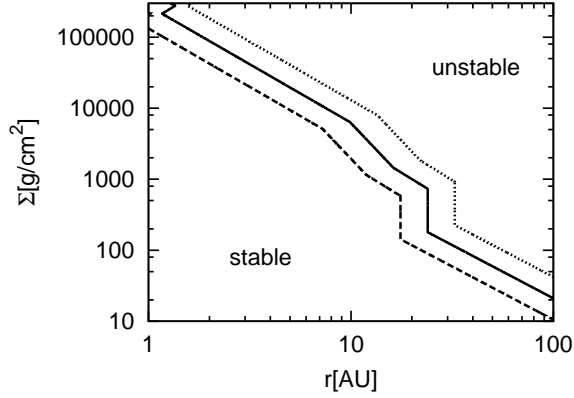}	
	\end{center}
	\caption{The critical surface density for the fiducial case. The solid line represents the critical surface density $\Sigma_c$ that is necessary to satisfy $Q=1$. Stable region is below this line and unstable region is above this line. The dotted and dashed lines represent the surface densities that is necessary to satisfy $Q=0.5$ and $Q=2$, respectively.}
		\label{fig:Qcntr}
\end{figure}

In figure \ref{fig:Qcntr}, $\Sigma_{\rm c}$ changes discontinuously by one order of magnitude at $r=24\ \rm{AU}$. 
We henceforth represent this critical radius as $r_{\rm c}$. 
By comparing figure \ref{fig:Tcntr} and figure \ref{fig:Qcntr}, it is confirmed that $ \Sigma_{\rm c} $ is large enough to satisfy $\tau\ge1$ in $r<r_{\rm c}$ because of the large Coriolis force and the high temperature. 
Analytic formula for $\Sigma_{\rm c}$ is derived from equations (\ref{ToomreQ}) and $\tau\ge1$ part of (\ref{temp}) as
\begin{equation}
\Sigma_{\rm c}=\left(\frac{27}{256}\kappa_0AB^aC^{6-2b+a}r^{-3(7+2a-2b)/2}\right)^{1/(4-2b)}\ \ \ (\tau >1)\  , \label{SigmaCrit}\\
\end{equation}
where $A$, $B$, and $C$ are defined in equations (\ref{A}), (\ref{B}), and 
\begin{equation}
C=\sqrt{\frac{\gamma k_{\rm B} M_{\rm s}}{\pi^2G\overline m}}\ ,\label{C}
\end{equation}
respectively. 
With the value of $\kappa_0$, $a$, and $b$ in table \ref{tab:BellLin}, equation (\ref{SigmaCrit}) provides us the property of $\Sigma_{\rm c}$ (see appendix \ref{app:TCSD} for details). 
In the fiducial case, in $1.2 \rm{AU}<r< 10\rm{AU}$, sublimation of metal dust mainly contributes to the opacity, and $\Sigma_{\rm c}\propto r^{-171/104}$. 
In $10 \rm{AU}<r< 16\rm{AU}$, the opacity is dominated by metal dust, and $\Sigma_{\rm c}\propto r^{-3}$. 
In $16 \rm{AU}<r< 24\rm{AU}$, the main component of opacity is sublimation of ices, and $\Sigma_{\rm c}\propto r^{-7/4}$. 
These properties can be seen in figure \ref{fig:Qcntr} from the fact that the $\Sigma_{\rm c}$ line breaks at the place where the main component of opacity changes. 
In $r>r_{\rm c}$, $\Sigma_{\rm c}$ is given by the isothermal condition $T=T_{\rm{min}}$ and $Q=1$ as 
\begin{equation}
\Sigma_{\rm c}=\frac{c_{\rm{s,min}}\Omega}{\pi G}=21\ {\rm g/cm}^2\ \left(\frac{r}{100{\rm AU}}\right)^{-3/2}\left(\frac{T_{\rm{min}}}{10{\rm K}}\right)^{1/2}\left(\frac{M_s}{\rm M_{\odot}}\right)^{1/2} \label{CSDiso}
\end{equation}
where $c_{\rm{s,min}}$ is the sound speed for $T=T_{\rm{min}}$.

In figure \ref{fig:Qcntr}, the critical surface density $ \Sigma_{\rm c} $ changes discontinuously at $ r = r_{\rm c} $. 
This is because the $ Q $ value is independent of the surface density there. 
The reason is explained as follows. 
If the surface density increases, the effect of self-gravity increases to destabilize a disk. 
On the other hand, in the optically thick case, the surface density also has an ability to block radiative cooling. 
Then, the effect of thermal pressure increases to stabilize a disk as the surface density increases. 
If these two effects are balanced, the value of Toomre Q becomes independent of the surface density. 
This balance is realized when the opacity depends on temperature as $\kappa\propto T^2$ in optically thick regime (see appendix \ref{app:TCSD}). 
The conditions to realize this balance are satisfied in the shaded area in figure \ref{fig:Tcntr} where ices become the main component of opacity in optically thick regime. 
Thus, the $\Sigma_{\rm c}$ discontinuously changes at the critical radius $r_{\rm c}$ where ices form. 
The analytical formula for critical radius $r_{\rm c}$ is derived by equations (\ref{ToomreQ}) and $\tau\ge1$ part of (\ref{temp}) with ices opacity as 
\begin{equation}
r_{\rm c}=\left(\frac{\gamma k_{\rm B} M_{\rm s}^{3/4}}{\pi \overline m G^{1/4}}\sqrt{\frac{27\kappa_0\alpha}{256\sigma}}\right)^{4/9}\simeq24{\rm AU}\left(\frac{\alpha}{0.01}\right)^{2/9}\left(\frac{M_{\rm s}}{\rm M_{{\odot}}}\right)^{1/3}\ . \label {rc}
\end{equation}
In equation (\ref{rc}), $r_{\rm c}$ is an increasing function of $\alpha$ and $M_{\rm s}$. 
This is because the viscous heating rate is large with large $\alpha$ and large $M_{\rm s}$. 
However, dependence on $\alpha$ and $M_{\rm s}$ is weak. 
Thus, $r_{\rm c}$ varies only severalfold even if $\alpha$ or $M_{\rm s}$ is different from typical value by an order of magnitude.

In figure \ref{fig:Qcntr}, it is seen that the $\Sigma_{\rm c}$ become multivalued function at $r\sim 1.5$AU and $\Sigma\sim2\times10^5\rm g/cm^2$. 
In this area, molecules are the main component of opacity, and there is the inner most radius of the unstable region around this area. 
We explain this inner most radius in appendix \ref{app:redge}. 

Note that the critical surface density is derived with the constant viscous parameter $\alpha$. 
It is considered that this assumption corresponds to the case that the viscosity originates from not the self-gravity but the MRI turbulence. 
This case assumes that the disk is MRI active everywhere. 
Thus, this result is specific to the case with the disk that has no region where the MRI is inactive (generally called "dead zones"). 
In other words, it is assumed that there is the significant heating of non-gravitational origin at all radii in the disk. 
On the other hand, the previous studies like \citet{Cla09} use the viscous parameter determined by the self-regulated gravitoturbulence with the condition $Q=1$. 
In this case, the disk is assumed to have the large dead zones without effective viscosity owing to the MRI turbulence. 
The realistic protoplanetary disks must be bracketed by these two extreme cases. 
Considering such disks is beyond the scope of this paper and is left as the future work.

\section{Instability Condition for Protoplanetary Disks} \label{sec:APP}
In this section, the condition for gravitational instability is discussed by comparing the surface density of particular disks to $\Sigma_{\rm c}$. 
We introduce two simple semi-analytic models for protoplanetary disks. 
One is the steady state accretion disk (the steady model). 
This model is realized if angular momentum is sufficiently transported. 
The other is the disk that has the same angular momentum distribution as the cloud core before it collapses. 
In this model, we assume the conservation of angular momentum distribution, and hence this model is hereafter called the AMC (Angular Momentum Conservation) model. 
The AMC model is realized if angular momentum is not sufficiently transported. 
We study the steady model in section \ref{sec:SSAD}. 
The AMC model is investigated in section \ref{sec:AMC}. 
The interpretation of these two models is discussed in section \ref{sec:int}.

\subsection{The Steady State Accretion Disk Model} \label{sec:SSAD} 

\begin{figure}[tbp]
	\begin{center}
\FigureFile(80mm,50mm){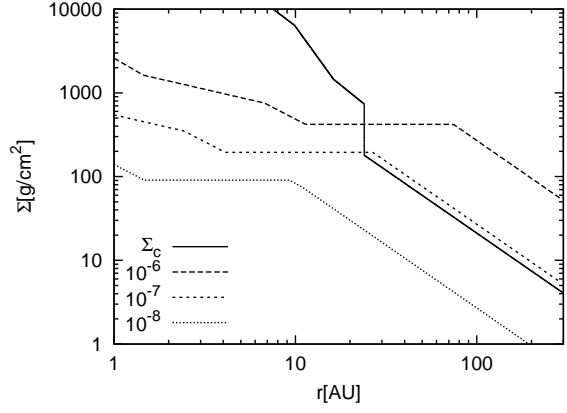}	
	\caption{Surface densities with $\dot M=10^{-6}$ (long-dashed line), $10^{-7}$ (short-dashed line), and $10^{-8} \rm M_{\odot}$ (dotted line). Solid line represent $\Sigma_{\rm c}$. We see the case of $\dot M=10^{-6}$ and $10^{-7}\rm M_{\odot}$ have the unstable region at $r>24$AU.}
		\label{fig:Steadydisk}
	\end{center}
\end{figure}
First, we consider the model of the steady state accretion disk. 
The surface density profile is calculated in the framework of the standard disk model with $\alpha$ viscosity (\cite{Sha73}; \cite{Pri81}). 
The disk structure are determined by three parameters; viscous parameter $\alpha$, protostar mass $M_{\rm s}$, and mass accretion rate $\dot M$. 
By the equation of continuity and angular momentum conservation, we have the relation
\begin{equation}
\alpha c_{\rm s} H \Sigma=\frac{\dot M}{3\pi}. \label{alphaMdot}
\end{equation}
This equation shows that surface density $\Sigma$ is determined by mass accretion rate $\dot M$. 
We calculate the temperature in the same manner as in section 2 using the cooling function approximated as equation (\ref{Q-}) and opacity represented by equation (\ref{BellLin}) with table \ref{tab:BellLin}, and introducing $T_{\rm min}$. 
Using equations (\ref{Omega}), (\ref{cs}), (\ref{H}), (\ref{temp}), and (\ref{alphaMdot}), we obtain the analytic profile for the surface density in this model as 
\begin{eqnarray}
\Sigma_{\rm{SD}}=
\left\{
\begin{array}{ll}
{\displaystyle D^{X}\left(\frac{27\kappa_0AB^a}{256}r^{-3(a+1)/2}\right)^{-\frac{1}{5+1.5a-b}}r^{-3X/2}} & (\tau \ge1)\\[10pt]
{\displaystyle D^{X}\left(\frac{27A}{256\kappa_0B^a}r^{-3(1-a)/2}\right)^{-\frac{1}{3+b-1.5a}}r^{-3X/2}} & (\tau <1)\\[10pt]
{\displaystyle \frac{\dot M\sqrt{GM_{\rm s}}}{3\pi\alpha c_{\rm{s,min}}^2}}r^{-3/2} & (T=T_{\rm{min}})
\end{array}
\right., \label{SigmaSD}
\end{eqnarray}
where we use a symbols $A,\ B,\ D$, and $X$ are defined in equations (\ref{A}), (\ref{B}),
\begin{eqnarray}
D&=&\frac{\overline m\dot M}{3\pi \alpha\gamma k_{\rm B}}\sqrt{GM_{\rm s}}\ , \nonumber\\
\end{eqnarray}
and
\begin{eqnarray}
X&=&
\left\{
\begin{array}{ll}
{\displaystyle \frac{3+a/2-b}{5+1.5a-b}} & (\tau \ge1) \nonumber \\[10pt]
{\displaystyle \frac{3+b-a/2}{3+b-1.5a}} & (\tau <1) \nonumber
\end{array}
\right. ,
\end{eqnarray}
respectively. 
This steady state accretion disk does not have the inner and the outer radius. 
In other words, the disk is infinitely extended. 

\begin{figure}[tbp]
	\begin{center}
\FigureFile(80mm,50mm){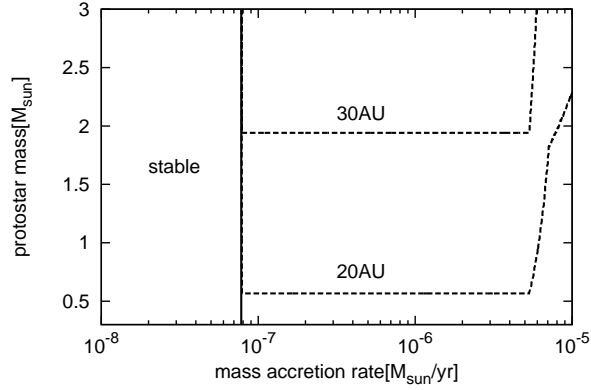}	
	\caption{The contours of minimum radius of unstable region $r_{\rm SD,min}$ for various values of $M_{\rm s}$ and $\dot M$. The dashed lines show contours of minimum radius $r_{\rm SD,min}$. The area where mass flux is less than $7.7\times 10^{-8}\rm {M_{{\odot}}/yr}$ (left of solid line) has a stable disk.}
		\label{fig:SDfrag}
	\end{center}
\end{figure}

From equation (\ref{SigmaSD}) and table \ref{tab:BellLin}, we can calculate the $\Sigma_{\rm SD}$. 
Figure \ref{fig:Steadydisk} shows the surface densities $\Sigma_{\rm SD}$ for the case with $\dot M=10^{-6}$ (long-dashed line), $10^{-7}$ (short-dashed line), and $10^{-8}\rm M_{\odot}/yr$ (dotted line) for the fiducial case defined in section \ref{sec:TEMP}. 
In figure \ref{fig:Steadydisk}, it is seen that the lines of $\Sigma_{\rm{SD}}$ break at the radii where the main component of opacity changes just like the $\Sigma_{\rm c}$ line in figure \ref{fig:Qcntr}. 
It is also seen that $\Sigma_{\rm{SD}}$ is large with large $\dot M$. 
The critical surface density $\Sigma_{\rm c}$ derived in section \ref{sec:CSD} is superimposed in figure \ref{fig:Steadydisk} by a solid line. 
The disk is unstable if the condition $\Sigma_{\rm{SD}}>\Sigma_{\rm c}$ is satisfied. 
From figure \ref{fig:Steadydisk}, it is found that the disks with $\dot M=10^{-6}$ and $10^{-7}\rm M_{\odot}/yr$ are unstable in $r>r_{\rm c}=24\rm AU$, where $r_{\rm c}$ is the critical radius defined in equation (\ref{rc}). 
On the other hand, the case with $\dot M=10^{-8}\rm M_{\odot}/yr$ is expected to be stable in all radius. 
It is also seen that all the lines including the $\Sigma_{\rm c}$ and $\Sigma_{\rm SD}$ has the same dependence on radius in the outer region ($r\gtrsim 100$AU). 
From equations (\ref{CSDiso}) and (\ref{SigmaSD}) for $T=T_{\rm min}$, we can confirm that $\Sigma_{\rm{SD}}$ and $\Sigma_{\rm c}$ has the same dependence on radius owing to isothermal state. 
By using this property, we can recast the instability condition $\Sigma_{\rm SD}\ge\Sigma_{\rm c}$ as
\begin{equation}
\dot M\ge \dot M_{\rm{crit}}\equiv \frac{3\alpha c_{\rm{s,min}}^3}{G}=7.7\times 10^{-8} {\rm M_{\odot}/yr}\left(\frac{\alpha}{0.01}\right)\left(\frac{T_{\rm min}}{10\rm K}\right)^{3/2}. \label{MdotC}
\end{equation}
Note that this instability condition (\ref{MdotC}) is independent of protostar mass $M_{\rm s}$. 
It is also found that $\dot M_{\rm{crit}}$ is increasing function of $\alpha$ and $T_{\rm min}$. 
As shown in figure \ref{fig:Steadydisk}, the minimum radius of the unstable region defined as $ r_{\rm SD,min} $ equals to the critical radius $ r_{\rm c} $. 
The critical radius $r_{\rm c}$ depends on the protostar mass as $r_{\rm c}\propto M_{\rm s}^{1/3}$, and $r_{\rm c}$ is independent of the mass accretion rate. 
Thus, in the steady model, mass accretion rate determines whether or not the disk becomes unstable, and protostar mass affects where the disk becomes unstable. 
This property described above is summarized in figure \ref{fig:SDfrag}. 
Figure \ref{fig:SDfrag} shows the contours of the minimum radius of the unstable region $ r_{\rm SD,min} $ in the $ \dot M - M_{\rm s} $ plane. 
It is seen that the stable disk is present in the left area where mass accretion rate is low and that the minimum radius is large with large protostar mass. 
It is also seen that the disk becomes unstable at smaller radius than $r_{\rm c}$ for the large mass accretion rate $\dot M\ge5.4\times10^{-6}\rm M_{\odot}/yr$. 
This property appears where the surface density $\Sigma_{\rm SD}$ is large enough to satisfy $\Sigma_{\rm SD}\ge\Sigma_{\rm c}$ in $r<r_{\rm c}$. 
From figure \ref{fig:Steadydisk}, it is found that $\Sigma_{\rm SD}$ is required to be larger than $7.3\times 10^2\rm g/cm^2$ in order to become gravitationally unstable in $r<r_{\rm c}$.

\subsection{The AMC Model} \label{sec:AMC}
Next, we consider the disk that has the same angular momentum distribution with the cloud core before it collapses. 
This model corresponds to the disk that has just formed. 

%%\subsubsection{The AMC model}
\begin{figure}[tbp]
	\begin{center}
\FigureFile(80mm,50mm){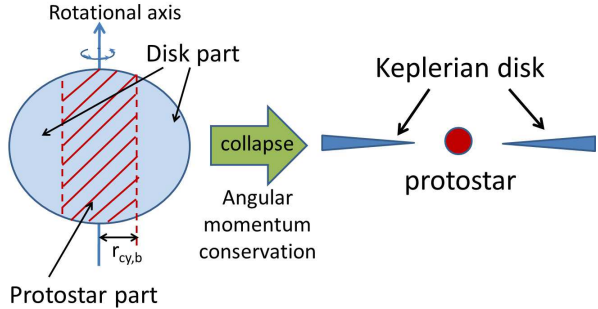}	
	\caption{Schematic representation of the AMC model. A rigidly rotating cloud collapses to make a protostar and a disk. The cloud core is divided into two parts. A protostar is made from the part near the rotational axis. A disk is made from the part that does not become protostar.}
		\label{fig:CloudCollapse}
	\end{center}
\end{figure}
Here, we introduce a simplified model for the formation of a disk and a protostar. 
Figure \ref{fig:CloudCollapse} schematically shows the formation process of the disk and the protostar in this model. 
It is assumed that a rotating cloud core collapses to make a protostar and a disk. 
We divide the cloud core into two parts. 
The shaded part near the rotation axis is assumed to become the protostar. 
The disk is assumed to be made from the other part. 
The boundary radius $r_{\rm cy,in}$ between these two parts is determined so that the mass inside the shaded part equals to a given protostar mass $M_{\rm s}$ as a parameter. 
The cloud core is assumed to rotate rigidly. 
The angular velocity of the cloud core $\Omega_0$ is determined by introducing the rotation parameter $\beta_0\equiv E_{\rm{rot}}/|E_{\rm{grav}}|$, where $E_{\rm rot}$ is the rotation energy and $E_{\rm grav}$ is the gravitational energy. 
Note that the disk formed in this model has the inner radius $r_{\rm d,in}$ and the outer radius $r_{\rm d,out}$, different from the steady model described in section \ref{sec:SSAD}. 
It is assumed that the protostar and the disk forms instantaneously and that the disk has the Keplerian rotation velocity with ignoring the self-gravity. 
We also assume the inviscid and axisymmetric formation of the disk. 
In other words, it is assumed that the relation between the mass and the angular momentum is kept without redistribution. 
As the former cloud core, two types of the density distributions are considered with the assumption of spherical symmetry. 
One is the Bonnor-Ebert sphere (\cite{Ebe55}; \cite{Bon56}), and the other is a sphere of uniform density. 
In this study, we consider both cases, and the result for the case with Bonnor-Ebert sphere is mainly described. 
We discuss the difference of these two cases later.

%%\subsubsection{The case with the Bonnor-Ebert sphere} \label{sec:BE}

%%Here, a Bonnor-Ebert sphere (BE sphere; \cite{Ebe55}; \cite{Bon56}) is considered as the former cloud core. 
The Bonnor-Ebert sphere is an isothermal sphere in which the thermal pressure balances the self-gravity and the external pressure. 
This sphere is identified by three quantities, the cloud radius $R_{\rm E}$, the sound speed $c_{\rm{s,0}}$, and the central density $\rho_{c}$. 
It is known that a Bonnor-Ebert sphere is unstable when the condition 
\begin{equation}
R_{\rm E}\ge R_{\rm{crit}}\equiv 6.46\frac{c_{\rm{s,0}}}{\sqrt{4\pi G\rho_{\rm c}}} 
\end{equation}
is satisfied. 
The Bonnor-Ebert sphere whose radius $R_{\rm E}$ equals to $R_{\rm{crit}}$ is called the critical Bonnor-Ebert sphere, and we use this critical Bonnor-Ebert sphere in this model. 
We set $c_{\rm{s,0}}=190\rm m/sec$, which corresponds to the typical sound speed in a molecular cloud. 
The central density is fixed as $\rho_{\rm c}=7.45 \times10^{-19} \rm g/cm^3$. 
With these parameters, the critical Bonnor-Ebert sphere has the mass $M_{\rm c}=1 \rm M_{\odot}$ and the radius $R_{\rm E}=1.0\times10^4$AU. 
Here, we approximate the density profile of the Bonnor-Ebert sphere (Tomide 2011, private communication) as 
\begin{equation}
\rho(R)=\frac{\rho_{\rm c}}{\left(1+R^2/R_{\rm c}^2\right)^{3/2}}\ ,\label{BEapp}
\end{equation}
where $R$ is the radius in the spherical coordinate and $R_{\rm c}=c_{\rm{s,0}}/\sqrt{1.5G\rho_{\rm c}}$. 
The equation (\ref{BEapp}) approximates the density profile of a Bonnor-Ebert sphere to the accuracy of the error within a few percent. 
This approximation enables us to calculate the surface density of the Bonnor-Ebert sphere $\Sigma_{\rm BE}(r_{\rm cy})$ analytically as
\begin{eqnarray}
\Sigma_{\rm BE}(r_{\rm cy})=2\int_{0}^{\left(R_{\rm E}^2-r_{\rm cy}^2\right)^{1/2}}\rho(R) dz=\frac{2\rho_{\rm c}}{1+r_{\rm cy}^2/R_{\rm c}^2}\sqrt{\frac{R_{\rm E}^2-r_{\rm cy}^2}{1+R_{\rm E}^2/R_{\rm c}^2}}, \label{BEsd0}
\end{eqnarray}
where $r_{\rm cy}$ is the radius in the cylindrical coordinate. 

In this model, the surface density of the disk $\Sigma_{\rm AMC}(r_{\rm d})$ is determined from the angular momentum distribution of the former cloud core as the follows. 
Once the rotation parameter $\beta_0$ is given, the angular momentum distribution of the cloud core is determined. 
Next, give the ratio of disk to protostar mass $M_{\rm d}/M_{\rm s}$ as a parameter, and the specific angular momentum distribution of the disk is determined. 
Suppose that a fluid element in the cloud core at cylindrical radius $ r_{\rm cy} $ falls onto the disk at cylindrical radius $ r_{\rm d} $. 
The disk is supported by centrifugal force and rotates Keplerian velocity at $ r_{\rm d} $. 
By the conservation of angular momentum, the relation between $r_{\rm cy}$ and $r_{\rm d}$ is found as
\begin{equation}
r_{\rm cy}=(GM_{\rm s}r_{\rm d}/\Omega_0^2)^{1/4}\ . \label{BER}
\end{equation}
By using the relation (\ref{BER}) with the relation of mass conservation
\begin{equation}
r_{\rm d}\Sigma_{\rm AMC}(r_{\rm d}) dr_{\rm d}=r_{\rm cy}\Sigma_{\rm BE}(r_{\rm cy})dr_{\rm cy}\ ,
\end{equation}
we derive the surface density of the disk $\Sigma_{\rm{AMC}}$ as 
\begin{equation}
\Sigma_{\rm{AMC}}(r_{\rm d})=\frac{1}{2}\frac{\rho_{\rm c}R_{\rm E}}{1+r_{\rm cy}^2/R_{\rm c}^2}\left(\frac{1-r_{\rm cy}^2/R_{\rm E}^2}{1+R_{\rm E}^2/R_{\rm c}^2}\right)^{1/2}\frac{r_{\rm cy}^2}{r_{\rm d}^2}\ . \label{BESD}
\end{equation}
Note that the $r_{\rm cy}$ depends on $r_{\rm d}$ as in the equation (\ref{BER}). 
From equation (\ref{BESD}), we can find that $\Sigma_{\rm AMC}$ depends on radius as $r_{\rm d}^{-2}$ in the region where $r_{\rm cy}>R_{\rm c}$. 
It is also found that $\Sigma_{\rm AMC}$ depends on radius as $r_{\rm d}^{-1.5}$ for $r_{\rm cy}<R_{\rm c}$. 
Thus, the enclosed mass in the disk does not diverge. 
In this study, we concentrate on the Keplerian rotating disk without self-gravity. 
In order to use this assumption with validity, the mass of the disk should be less than that of the protostar. 
If we request that the disk rotates Keplerian velocity, the boundary radius $ r_{\rm cy,in} $ shown in figure \ref{fig:CloudCollapse} is always larger than $ R_{\rm c} $. 
Thus, the $\Sigma_{\rm AMC}$ depends on radius as $r_{\rm d}^{-2}$ in this paper. 
By the way, realistically, it is considered that a protostar does not have such a large mass just after the formation of the protoplanetary disk. 
In this model, we consider only the outer region of the disk and regard the inner region of the disk as a part of the protostar, at least, in the sense of gravitational field. 
The boundary radius $r_{\rm d,in}$ between the inner region and the outer region of the disk is determined by the equation (24) and $r_{\rm cy,in}$.

%%\subsubsection{result of outer disk model}
\begin{figure}[tbp]
	\begin{center}
\FigureFile(80mm,50mm){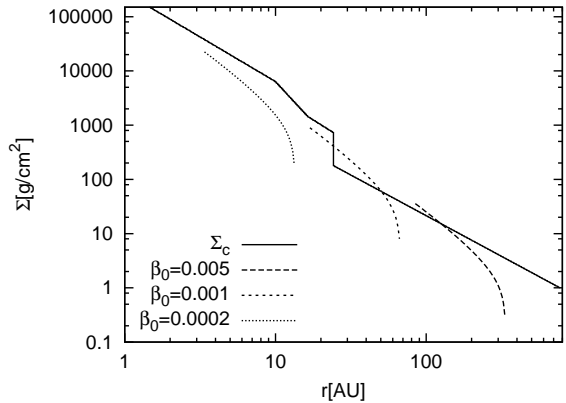}
	\caption{Surface density with $\beta_0=0.005$ (long-dashed), 0.001 (short-dashed), 0.0002 (dotted) and $M_{\rm d}/M_{\rm s}=0.25$. The critical surface density $\Sigma_{\rm c}$ is superimposed by solid line.}
		\label{fig:besd}
	\end{center}
\end{figure}
Now, we can calculate the surface density of the disk for given $M_{\rm d}/M_{\rm s}$ and $\beta_0$. 
First, we look over the property of $\Sigma_{\rm AMC}$ for the case with fixed $M_{\rm d}/M_{\rm s}$. 
Figure \ref{fig:besd} shows the surface densities $\Sigma_{\rm{AMC}}$ for the cases with $\beta_0=$0.005 (long-dashed), 0.001 (short-dashed), and 0.0002 (dotted) with $M_{\rm d}/M_{\rm s}=0.25$. 
In the case with large $ \beta_0 $, it is seen that $ \Sigma_{\rm{AMC}} $ is small and that the disk is present far from the protostar. 
The solid line in figure \ref{fig:besd} represents the critical surface density $\Sigma_{\rm c}$ derived in section \ref{sec:CSD}. 
In the case with $\beta_0=0.0002$, the disk is stable because the stability condition $\Sigma_{\rm c}>\Sigma_{\rm AMC}$ is satisfied owing to large $\Sigma_{\rm c}$ in $r<r_{\rm c}=22$AU, where $r_{\rm c}$ is the critical radius defined in equation (\ref{rc}) with $M_{\rm s}=0.8\rm M_{\odot}$. 
In the case with $\beta_0=0.001$, the stability condition $\Sigma_{\rm c} > \Sigma_{\rm AMC}$ is violated in $r_{\rm c}=22{\rm AU}\le r\le 54$AU, and thus the disk is unstable there. 
This is because $\Sigma_{\rm c}$ changes discontinuously at $r=r_{\rm c}$. 
In the case with $\beta_0=0.005$, the angular momentum is so large that the inner radius $r_{\rm in}=85$AU is larger than the critical radius $r_{\rm c}=22$AU. 
The disk can be unstable around the inner radius in $r_{\rm in}=85 {\rm AU}\le r\le 1.5\times10^2$AU.
Note that, in the cases with $\beta_0=0.001$ and 0.005, the outer regions of the disks are stable because $\Sigma_{\rm AMC}$ rapidly decreases as radius increases. 

%%This is because the disk with the large $\beta_0$ is present far from the protostar due to the large angular momentum of the former cloud core. 

\begin{figure}[tbp]
	\begin{center}
\FigureFile(80mm,50mm){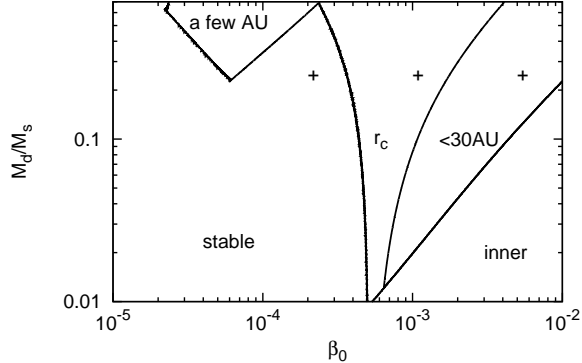}
	\caption{The unstable condition on $\beta_0$-$M_{\rm d}/M_{\rm s}$ plane for the case with Bonnor-Ebert sphere. The areas labeled as ``stable" and ``inner" represents a stable disk in the AMC model. The areas labeled as ``a few AU", ``r$_{\rm c}$", and ``$<$30AU" have an unstable disk. ``+" symbols denote the cases of figure \ref{fig:besd}. See text for details.}
		\label{fig:befragcntr}
	\end{center}
\end{figure}
Next, we calculate $\Sigma_{\rm AMC}$ for various values of $M_{\rm d}/M_{\rm s}$ as well as $\beta_0$. 
Figure \ref{fig:befragcntr} summarizes the results in $\beta_0$-$M_{\rm d}/M_{\rm s}$ plane. 
The symbols with ``+"  correspond to the cases in figure \ref{fig:besd}. 
The results are classified into five groups in the $\beta_0-M_{\rm d}/M_{\rm s}$ plane. 
The area labeled as ``stable" in figure \ref{fig:befragcntr} indicates that the disks with these parameters are stable in all radius. 
The case with $\beta_0=0.0002$ in figure \ref{fig:besd} belongs to this area. 
These disks have small outer radius $r_{\rm out}<r_{\rm c}$ owing to the small angular momentum. 
Thus, the condition $\Sigma_{\rm AMC}<\Sigma{\rm c}$ is satisfied due to the large $\Sigma_{\rm c}$ in $r<r_{\rm c}$. 
In figure \ref{fig:befragcntr}, the areas labeled as ``a few AU", ``r$_{\rm c}$", and ``$<$30AU" indicate the unstable disks. 
The area labeled as ``a few AU" exists in the small $\beta_0$ and large $M_{\rm d}/M_{\rm s}$ area. 
The disks with these parameters have small radius and large surface density enough to satisfy $\Sigma_{\rm AMC}>\Sigma_{\rm c}$ in $r<r_{\rm c}$. 
Thus, they have a possibility of fragmentation at a few AU from the protostar. 
The area labeled as ``r$_{\rm c}$" exists around middle $\beta_0$ area in figure \ref{fig:befragcntr}. 
This area includes the case with $\beta_0=0.001$ in figure \ref{fig:besd}. 
Although the disks with this area are stable inside the critical radius $r_{\rm c}$ due to small $\Sigma_{\rm AMC}$, they are unstable in $r\ge r_{\rm c}$. 
This is because $\Sigma_{\rm c}$ drastically changes at $r=r_{\rm c}$. 
It is the formation of ices that enables the disks to become unstable. 
The area labeled as ``$<$30AU" exists at large $\beta_0$ area. 
This area includes the case with $\beta_0=0.005$ in figure \ref{fig:besd}. 
In this area, the inner radius $r_{\rm in}$ of the disk is larger than the critical radius $r_{\rm c}$, and the disk is unstable around the inner radius. 
The area labeled as ``inner" in figure \ref{fig:befragcntr} is difficult to conclude. 
In the present AMC model, the disks with this area are regarded as stable because $\Sigma_{\rm AMC}$ is very small due to the too large inner radius $r_{\rm in}\gtrsim 10^2$ AU. 
However, this large $r_{\rm in}$ is artificial because the AMC model is too simplified in order to calculate analytically. 
Realistically, it is expected that the disk is present in $r<r_{\rm in}$. 
If the disk has the surface density represented as in equation (\ref{BESD}), the disks with the area labeled as ``inner" in figure \ref{fig:befragcntr} have a possibility to be unstable. 

From above results, we found that the disk is expected to be unstable with large $\beta_0$ in the AMC model. 
Here, we estimate the critical rotation parameter $\beta_{\rm 0,c}$ that is necessary to become unstable for the disk with small mass ($\sim0.01M_{\rm s}$). 
In this model, the disk becomes unstable if the disk extends to the radius $r\ge r_{\rm c}$. 
The outer radius of the disk can be estimated by substituting $r_{\rm cy}=R_{\rm E}$ in equation (\ref{BER}). 
On the other hand, the $r_{\rm c}$ is represented in equation (\ref{rc}). 
By using the relation $\beta_0\propto \Omega_0^2$, we derive the critical rotation parameter $\beta_{\rm 0,c}$ as 
\begin{equation}
\beta_{\rm 0,c}=4\times10^{-4}\left(\frac{r_{\rm c}}{24\rm AU}\right)^{}\left(\frac{R_{\rm E}}{10^4\rm AU}\right)^{4}\left(\frac{M_{\rm c}}{1 \rm M_{\odot}}\right)^{}.
\end{equation}
Note that the $r_{\rm c}$ depends on the viscous parameter $\alpha$ and on the protostar mass $M_{\rm s}$ and that the $R_{\rm E}$ and $M_{\rm c}$ depend on the central density $\rho_{\rm c}$ and sound speed $c_{\rm s,0}$. 
In figure \ref{fig:befragcntr}, it is seen that the disk becomes unstable for the $\beta_0\gtrsim\beta_{\rm 0,c}$ even with small mass ratio $M_{\rm d}/M_{\rm s}\lesssim 0.1$. 
The minimum ratio of disk to protostar mass satisfying the unstable condition is $M_{\rm d}/M_{\rm s}|_{\rm min}\simeq0.0092$ with $\beta_0=5.0\times10^{-4}$, which belongs to ``r$_{\rm c}$" area. 
%%Therefore, in this model, the disks are likely to become unstable at the critical radius $r_{\rm c}$. 

\begin{figure}[tbp]
	\begin{center}
\FigureFile(80mm,50mm){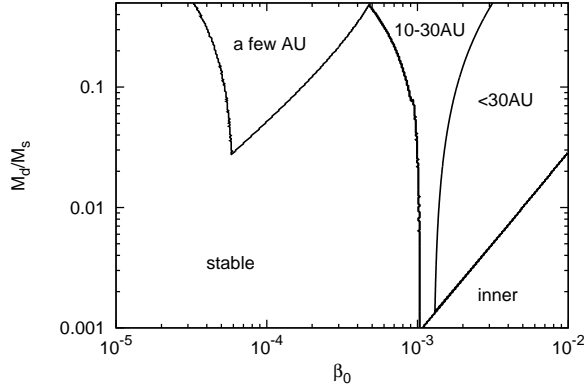}
	\caption{The unstable condition on $\beta_0$-$M_{\rm d}/M_{\rm s}$ plane for the case with the uniform sphere. The areas labeled as ``stable" and ``inner" represents a stable disk. The disks with the parameter areas labeled as ``a few AU", ``10-30AU", and ``$<$ 30 AU" become unstable at the labeled radius.}
		\label{fig:udfc}
	\end{center}
\end{figure}

We also perform the calculation for the case with the uniform-density sphere as the density distribution of the former core. 
In this case, the core is assumed to rotate rigidly in common with the case with the Bonnor-Ebert sphere. 
This uniform sphere is characterized by two parameters, $\alpha_0\equiv E_{\rm th}/|E_{\rm grav}|$ and $\beta_0\equiv E_{\rm rot}/|E_{\rm grav}|$, where $\alpha_0$ is the thermal parameter and $E_{\rm th}$ is the thermal energy. 
The thermal parameter is assumed to be $ \alpha_0 = 0.86 $, which is the same value as the case with the critical Bonnor-Ebert sphere. 
The calculation is performed with various values of $\beta_0$ and $M_{\rm d}/M_{\rm s}$. 
Figure \ref{fig:udfc} summarizes the results. 
The results are qualitatively same as the case with the Bonnor-Ebert sphere but quantitatively different at some points. 
One is that the critical rotation parameter $\beta_{\rm 0,c}$. 
For the case with the uniform sphere, the value $\beta_{\rm 0,c}= 1.0\times10^{-3}$ is larger than that with the Bonnor-Ebert sphere by factor 2.5. 
This difference arises from the fact that the uniform-density sphere has 1.2 times smaller radius $R_{\rm E}$ and almost same angular velocity $\Omega_0$ as compared to those of the critical Bonnor-Ebert sphere with the same values of $\alpha_0$ and $\beta_0$. 
Since a sphere with a small radius makes a disk that has a small outer radius with a fixed angular velocity, the case with the uniform sphere needs larger $ \beta_0 $ than that with the Bonnor-Ebert sphere in order to satisfy the condition $ r_{\rm out} \ge r_{\rm c}$. 
Next, for the same radius $r$ in the disk, the mass ratio $M_{\rm d}/M_{\rm s}$ that is necessary for the instability is smaller with the uniform sphere than with the Bonnor-Ebert sphere by nearly one order of magnitude. 
In the AMC model, the disk is made from the outer part of the sphere, and the density of the outer part with the uniform sphere is larger than that with the Bonnor-Ebert sphere. 
Thus, if both spheres make the disks in same size, the disk made from the uniform sphere has a larger surface density than that from the Bonnor-Ebert sphere. 
Hence, for the same radius $r$, the disk made from the uniform sphere needs smaller $M_{\rm d}/M_{\rm s}$ to become unstable than that from the Bonnor-Ebert sphere. 
Finally, the instability radius around the mid $\beta_0$ is different. 
In the case with the Bonnor-Ebert sphere, the instability radius is clearly divided into two parts labeled as ``a few AU" and ``r$_{\rm c}$". 
On the other hand, the case with the uniform sphere, the disk becomes unstable at the intermediate radius 10AU-30AU. 
This is due to the large surface density and its strong dependence on the radius of the disk from the uniform sphere. 
All of these differences between the disk from the Bonnor-Ebert sphere and from the uniform sphere originate from the difference in the relation between mass and specific angular momentum (known as $m-j$ relation). 
Moreover, these differences strongly depend on the assumption of rigid rotation. 

\if0
The minimum mass ratio $M_{\rm d}/M_{\rm s}|_{\rm min}$ for the case with uniform sphere is smaller than that with Bonnor-Ebert sphere. 
With the uniform sphere, the surface density of the disk more rapidly decrease as radius increase than that with Bonnor-Ebert sphere. 
Thus, the disk 
%%This is because the angular momentum is so large that $\Sigma_{\rm AMC}$ cannot attain the value of $\Sigma_{\rm c}$. 
%%However, it is considered that protoplanetary disks may not have such a large inner radius except the transitional disks (refは探しておきます). 
%%because the AMC model corresponds to the disk that has just formed (discussed in next subsection). 
With the same $\beta_0$, the value of angular velosity $\Omega_0$ with these two spheres are almost the same. 
If $\Omega_0$ is the same, the outer radius of the disk is large with large sphere radius. 
Hence, the outer radius of the disk with the Bonnor-Ebert sphere case attains the critical radius $r_{\rm c}$ with smaller $beta_0$ than in the uniform density sphere case. 
Since the disk with the case of uniform density sphere has small radius, the surface density with the case of uniform density sphere become larger than that with the Bonnor-Ebert sphere case. 
Thus, the area size labeled as ``grains" become large in the case of the uniform density sphere.
\fi

\subsection{Interpretation of the Steady Model and the AMC Model} \label{sec:int}

In previous two subsections, we studied the gravitational stability of protoplanetary disks by using two simple semi-analytic models. 
Here, we interpret these models by considering the formation scenario of a protoplanetary disk. 

%%The scenario of the disk formation is as follows.
%%The formation process of a protostar and a protoplanetary disk is devided into two phase; the ranaway collapse phase and the inside-out collapse phase. 
%%The earlier phase is called the ranaway collapse, and the later phase is called the inside-out collapse. 

It is believed that a cloud core with angular momentum collapses to form a protostar and a protoplanetary disk. 
First, the central part of the cloud core begins to contract and makes a protostar and a small (typically a few AU) disk (e.g. \cite{Bat98}). 
Next, the outer envelope of the core mainly falls onto the disk, and the mass accretion from the envelope makes the disk grow up in mass and size as typically larger than 100 AU (e.g. \cite{Ada88}). 
The time scale for this disk growth is estimated as $t_{\rm{grow}}\sim r/c_{\rm{s,0}}$ according to the self-similar solution for the collapse of the rotating isothermal cloud \citep{Sai98}. 
This disk that has just formed corresponds to the AMC model introduced in section \ref{sec:AMC}. 
Later, this disk experiences the viscous evolution with the transport of angular momentum and eventually forgets the initial distribution of the angular momentum. 
The time scale of viscous evolution is estimated as $t_{\rm{vis}}\sim \alpha^{-1}(H/r)^{-2}\Omega^{-1}$ from the azimuthal component of the momentum equation. 
It is expected that a steady state disk, which is discussed in section \ref{sec:SSAD}, corresponds to the state after the viscous evolution. 
For the case with $c_{\rm s,0}=190\rm m/sec$ and $H/r=0.1$, the time scales for disk growth $t_{\rm grow}$ and for viscous evolution $t_{\rm vis}$ are estimated as 
\begin{equation}
t_{\rm{vis}}\sim 1.6\times 10^{3} {\rm yr}\left(\frac{\alpha}{0.01}\right)^{-1}\left(\frac{r}{1\rm{AU}}\right)^{3/2}\label{tvis} \\
\end{equation}
and
\begin{equation}
t_{\rm{grow}}\sim 25 {\rm yr}\left(\frac{c_{\rm{s,0}}}{190\rm m/s}\right)^{-1}\left(\frac{r}{1\rm{AU}}\right)^{1}. \label{tgrow}
\end{equation}
From equations (\ref{tvis}) and (\ref{tgrow}), it is seen that the disk growth time is much shorter than the viscous evolution time for $r>1$AU. 
Thus, it is expected that the viscous redistribution of angular momentum can be regarded as negligible during much longer time than the dynamical time scale $\Omega^{-1}$ after the formation of the disk. 
From equation (\ref{tvis}), it is found that the viscous evolution time is short for a small radius. 
Hence, it is likely that the steady state accretion disk forms at inner region first, and it will spread outward in viscous time scale. 
In summary, we interpret the AMC model as the outer part of the young disk and the steady model as the old disk.

\section{Discussion}\label{sec:DIS}
%%\subsection{the critical surface density}
%%Here, we consider the validity of the critical surface density derived in section \ref{sec:CSD}. 

\subsection{Thermal Stability of Equilibrium State}

We calculated the equilibrium temperature based on the approximated opacity given in equation (\ref{BellLin}) and table \ref{tab:BellLin} \citep{Bel94} in section \ref{sec:TEMP}. 
This opacity depends on the temperature, and the property of the opacity varies when the main component of opacity changes. 
Thus, the cooling rate also depends on the temperature similarly. %%, and the property of it varies with opacity. 
If the temperature dependence of the cooling rate is weaker than that of the heating rate, this state is thermally unstable \citep{Fie65}. 
In order to check the stability of the thermal equilibrium state we used, we compare the exponent of the temperature in the heating rate to the exponent in the cooling rate. 
In our model, as shown in section \ref{sec:TEMP}, the heating rate has the relation $\Gamma\propto T$, regardless of the optical depth. 

In optically thick case, the temperature dependence of the cooling rate is $\Lambda\propto T^{4-b+a/2}$. 
Thus, the condition for thermal stability is represented as $a-2b>-6$. 
This condition is satisfied except for the case that hydrogen scattering is the main component of the opacity. 
This case is realized where the temperature are larger than the critical temperature $T\ge T_{\rm TI}\sim3300$K for $\rho=10^{-10}\rm g/cm^2$. 
Within the parameter range we are interested in, the equilibrium temperature is safely smaller than $T_{\rm TI}$, and thus, in optically thick case, the equilibrium state is stable. 

In optically thin case, the temperature dependence of the cooling rate is $\Lambda\propto T^{4+b-a/2}$. 
The condition for thermal stability is obtained as $a-2b<6$. 
From the values in table \ref{tab:BellLin}, this stability condition is violated when the main component of the opacity is sublimation of ices ($T_1=166.8{\rm K}<T<202.6\rm K$) or sublimation of metal dust ($2286.7\rho^{2/49}{\rm K}<T<2029.7\rho^{1/81}$). 
According to results in section \ref{sec:TEMP}, in the optically thin case, the equilibrium temperature is sufficiently smaller than $T_1$ above which sublimation of ices becomes the main component of the opacity. 
Thus, in optically thin case, the equilibrium state satisfies the condition of thermal stability. 
Therefore, the equilibrium state we used in this study is thermally stable. 
%%Thus, there is a possibility that the heating rate increases faster than cooling rate when the temperature increases. 
%%If this condition is satisfied, the equilibrium temperature is not realized because  thermal nstable . 
%%we derive the thermal instability condition. 

\subsection{Comparison to Radiation Hydrodynamical Simulations}
Here, we compare the results of our model to the results of some radiation hydrodynamical (RHD) simulations. 
First example is the simulation by \citet{Bol06}. 
They calculated the disk evolution about 4000 yr, and the surface density of the disk at the final state was approximated as $\Sigma\simeq 150{\rm g/cm^2} \times 10^{-(r/R_{\rm E})^2}$, where $R_{\rm E}=46.7{\rm AU}$. 
This disk did not fragment in their simulation. 
Our model also predicts that this disk cannot fragment because this surface density satisfies the stable condition $\Sigma<\Sigma_{\rm c}$. 
Thus, our model is consistent with the result of their simulation.

Next, we make the comparison to the result by \citet{Mer10}. 
In their study, the evolutions of disks were calculated for the case with different values of the opacities. 
Their result showed that the opacity that is smaller than the interstellar values promotes the fragmentation. 
In our model, the values of $\Sigma_{\rm c}$ and $r_{\rm c}$ represented in equations (\ref{SigmaCrit}) and (\ref{rc}) are small with the small opacity. 
This means that the disk easily becomes unstable with small opacity. 
In this sense, the prediction by our model is consistent with them. 

Finally, we compare the result of the simulation by \citet{Sta11}. 
Their calculation was performed with nine initial conditions, and they obtained the result that only three of nine disks fragmented. 
From this result, they suggested that the fragmentation is unlikely to occur when the disk satisfies the condition that $r_{\rm out}< 100$AU and that $M_{\rm d}/M_{\rm s}<$0.36, where $r_{\rm out}$ is the outer radius of the disk and $M_{\rm d}/M_{\rm s}$ is the ratio of disk to protostar mass. 
Our result is consistent with their result in the sense that fragmentation is difficult near the central star ($r\lesssim 20$AU). 
However, with their initial surface density profile, our model predicts that all of nine disks used in \citet{Sta11} become gravitationally unstable for the case with $ \alpha \lesssim0.6 $. 
These differences between their results and our predictions may be caused by the heating due to the gravitoturbulence. 
In this sense, we should notice that the instability condition investigated in this paper is not exactly the same as the fragmentation condition.

%%From above two comparison, it is thought that our model has the ability to re-create the results of 3D RHD simulations. 

%%However, there is some simulations whose results is not consistent with our model. %%(e.g. \cite{Mer10}; ). \citet{Stam11} 

\if0
It seems that our model is inconsistent with their calculation, but 
They initially set the mass of a central star $M_{\rm s} = 0.7\rm M_{\odot}$, the surface density $\Sigma_0 \propto r^{-1}$, and temperature $T\propto r^{-q}$. 
%%We choose 3D simulations without effect of irradiation from the central star as targets of comparison because we neglect irradiation.
%%By drawing the surface density of this initial condition on the $r-\Sigma$ plane, we confirm this disk is present in the unstable region. 
%%We make two comparisons. 
They initially set mass of a central star $M_{\rm s} = 0.7\rm M_{\odot}$, disk mass $M_{\rm d}=0.7\rm M_{\odot}$, inner radius $R_{\rm in}=40\rm{AU}$, outer radius $R_{\rm out}=400\rm{AU}$, surface density $\Sigma_0 \propto r^{-7/4}$, and temperature $T\propto r^{-1/2}$. 
The numerical result showed that the disk fragmented and that the 7 objects are produced by fragmentation. 
Our model also predicts that this disk will fragment because this initial surface density is larger than the critical surface density $\Sigma_{\rm c}$ introduced in section \ref{sec:CSD}. 
Hence, our model is consistent with the result of their simulation. 
\fi

\subsection{The Steady Disk Model}

%%Hartmann1998と比べて具体的な数字書け

In section \ref{sec:SSAD}, we derived the critical mass accretion rate $\dot M_{\rm crit}$ that is necessary in order that the steady disk becomes unstable. 
\citet{Har98} derived the mass accretion rate of T-Tauri stars in Taurus and Chamaeleon I, based on the optical observation. 
The typical value of the accretion rate is as small as $\dot M\sim 10^{-8}\rm M_{\odot}/yr$. 
Since this value is less than the critical mass accretion rate $\dot M_{\rm crit}\simeq 7.7\times10^{-8}\rm M_{\odot}/yr$ given in equation (\ref{MdotC}), the disks around a T-Tauri stars are expected to be stable. 
However, in \citet{Har98}, 7 of 40 T-Tauri stars in Taurus and 1 of 16 in Chamaeleon I are estimated to have a large mass accretion rate enough to become gravitationally unstable. 
Thus, these statistics indicate that about 18 \% of the T-Tauri stars have the possibility to have an unstable disk in Taurus, whereas the fraction is less than 7 \% in Chamaeleon I. 
By the way, around a protostar, the Keplerian disk has not been clearly observed yet. 
If such a disk is present, its mass accretion rate is estimated as 
\begin{equation}
\dot M\sim \frac{M_{\rm J}}{t_{\rm{ff}}}\sim\frac{c_{\rm s}^3}{G}\sim2\times10^{-6}\rm M_{\odot}/yr\left(\frac{T}{10\rm K}\right)^{3/2}, \label{PSMdot}
\end{equation}
where $M_{\rm J}$ is Jeans mass, and $t_{\rm ff}$ is free-fall time. 
Because this mass accretion rate is much larger than $\dot M_{\rm crit}$, the disk around a protostar is expected to be gravitationally unstable. 
In summary, it is expected that most of the disk around a T-Tauri star is stable and that the disk around a protostar is likely to fragment. 
Anyway, the relation between mass accretion rate and gravitational instability may be a useful tool to probe the fragmentation in the disk systems observationally. 

\citet{Cla09} (hereafter C09) also discussed the fragmentation of a steady state accretion disk. 
Assumptions set in C09 are similar to our model, but there are critical differences. 
C09 assumes the condition of gravitoturbulence $Q=1$, uses spatially-variable $\alpha$ parameter, and adopts Gammie criterion as fragmentation criterion. 
On the other hand, we assume not to be in gravitoturbulent state, use constant viscous parameter $  \alpha$, and adopt Toomre criterion. 
These differences produce qualitatively different results. 
For example, in C09, fragmentation is predicted even with very small mass accretion rate ($\dot M\sim10^{-8}\rm M_{\odot}/yr$), whereas, in our model, the disk is expected to be stable with such a small mass accretion rate. 
However, note that both studies imply that the formation of ices is an important process. 
In the regime of ices opacity $\kappa\propto T^2$, the temperature dependence of cooling rate becomes $\Lambda\propto T^2$, which is weaker than that of other opacities (sublimation of ices, metal dust, and sublimation of metal dust). 
This property enables the disk to become low temperature when the heating rate becomes small. 
The importance of this fact does not change regardless of fragmentation criteria.

%%Although these two studies use different criteria to determine whether the disk can fragment, 
%%Therefore, we stress that it is important to consider a realistic opacity when discussing the possibility and position of the fragmentation.
%%The disks around these T-Tauri stars have the possibility to fragment or have spiral arms. 
%%In the core accretion model (\cite{Gol73}; \cite{Hay81}; \cite{Pol96}), it is considered that ices is necessary to form giant planet because ices occupy the large fraction of the mass of dust grains. 

\subsection{The AMC Model}
%%Here, we discuss the AMC model introduced in section \ref{sec:AMC}. 
%%Here, we compare the result of the previous calculation of the core collapse to the result of the AMC model introduced in section \ref{sec:AMC}. 

In the AMC model introduced in section \ref{sec:AMC}, the disk is unstable when the condition $\beta_0\ge \beta_{\rm 0,c}\sim 0.001$ is satisfied. 
By the way, according to the results of radiation hydrodynamical calculations, the disk made from the core with the large rotation parameter $ \beta_0 \ge \beta_{\rm 0,b} \sim 0.01$ is expected to fragment before the formation of the protostar (\cite{Bat11}). 
From the observation of the line emission of NH$_3$ and N$_2$H$^+$, the rotation parameter $\beta_0$ of the cloud core is estimated to range in $10^{-6}< \beta_0<0.07$ with the typical value $\beta_0\sim 0.02$ \citep{Cas02}. 
Only the 3 of 20 cores have the rotation parameter $\beta_0\le \beta_{\rm 0,c}\sim0.001$. 
Thus, only 15 \% of the cores are expected to make a stable disk. 
The cores with middle rotation parameter $\beta_{\rm 0,c}\sim 0.001\le \beta_0\le \beta_{\rm 0,b}\sim0.01$ have the possibility to fragment after the formation of a protostar. 
The 8 of 20 cores is present within this parameter range. 
The number of the core with $\beta_0\ge\beta_{\rm 0,b}\sim0.01 $ is the 9 of 20 cores. 
The cases with such a large $\beta_0$ are outside the scope of our AMC model. %% because the disks fragment before the formation of the protostar. 

%%コア収縮のシミュレーションと比較しましょう

\citet{Mat03}, \citet{Mac10}, and \citet{Tsu11} calculated the formation process of the disk from the cloud core using barotropic equation of state for various values of the rotation parameter $\beta_0$. 
These studies showed that the disks become gravitationally unstable for the cases with large $\beta_0$. 
This feature is consistent with the result of our AMC model. 
\citet{Mat03} calculated the cloud collapse with the rotation parameter $8\times10^{-4}\le\beta_0\le8\times10^{-2}$. 
They showed that the disk made from the cloud with $\beta_0\simeq8\times10^{-4}$ become gravitationally unstable but not fragmenting until the final state of their calculation. 
Our AMC model also predicts that the disk with $\beta_0\simeq8\times10^{-4}$ is unstable. 
They also showed that fragmentation occurred before the formation of the protostar for the case with $\beta_0\gtrsim2\times10^{-3}$. 
These cases are outside the scope of our AMC model. 
The result in \citet{Mac10} is the quantitatively different from our model. 
For example, the value $\beta_{0,c} \sim 3 \times 10^{-5} $ in their calculation is smaller than that in the AMC model by one order of magnitude. 
However, it is difficult to compare the result of the AMC model to that of their calculation quantitatively because, in their calculation, fragmentation tends to occur during the phase that the mass of protostar is smaller than that of the disk. 
\citet{Tsu11} is more consistent with our AMC model. 
Despite that they used the barotropic equation of state, which is different from us, the value $\beta_{\rm 0,c}=3\times10^{-3}$ in their calculation is consistent with our AMC model within the error of factor 3. 
%%Here, we define $\beta_{\rm 0,c}$ in TM11 as the rotation parameter that is necessary to form gravitationally unstable disk. 

\subsection{The effects of ignored processes}
In this study, we ignored some physical processes that are expected to affect the gravitational instability in the disk. 
Here, we comment on the effects of processes ignored in this paper. 

First, the self-gravity will affect the disk properties through the effect on the transport of angular momentum and energy dissipation. 
In the form of gravitational torque due to the non-axisymmetric mode and the gravitoturbulence, the self-gravity affects the viscous parameter $\alpha$. 
It is discussed that the $\alpha$ becomes large if the $Q$ value approaches unity \citep{Kra08}. 
However, this $\alpha$ as a function of disk properties has not been clarified yet. 
In this study, in order to avoid this uncertainty, we used constant $\alpha$ parameter and derived the condition for gravitational instability as a function of $\alpha$. 
The work considering $\alpha$ from the self-gravity remains as a future work. 
The self-gravity will also affect the scale height $H$. 
If considering the self-gravity in the vertical direction, the scale height $H_{\rm sg}$ with self-gravity is represented as a function of $Q$ value. 
This $ H_{\rm sg} $ for the case with $ Q = 1 $ is approximately half of the $H$ without self-gravity represented in equation (\ref{H}). 
From this effect, the values of $\Sigma_{\rm c}$ and $r_{\rm c}$ are modified, but the amount of this modification is at most a few tens of percent. 
Thus, it is considered that our result does not qualitatively change even if considering the compression by self-gravity in the vertical direction. 

Next, the effect of irradiation from the protostar will affect the disk properties. 
At the region far from the protostar, the heating by irradiation from the protostar becomes greater than the viscous heating because the viscous heating strongly decreases as the radius increases. 
Thus, at the far from the protostar, there is a possibility that the disk is stabilized by the irradiation heating. 
The heating rate by irradiation is affected by the disk flaring, and this flaring is determined by the temperature distribution of the disk. 
It is difficult to make the accurate treatment of irradiation heating in our model because we did not consider the radial distribution of the disk when deriving the instability condition in section \ref{sec:CSD}. 
However, the effect of irradiation heating is weak in disks with large surface density. 
Thus, it is considered that our results for optically thick case will not critically change. 
This irradiation heating should be taken into account when investigating the properties of the particular disk.

Finally, we comment on the effect of magnetic field. 
The magnetic field has much influence on the angular momentum transport during all phases from the protostar formation to the disk evolution. 
In the phase of the protostar formation, it is considered that the magnetic field triggers the outflow and that it transports the mass and the angular momentum (\cite{Shu94}; \cite{Tom02}; \cite{Mac08}). 
Thus, it is expected that the assumption used in the AMC model is violated by the effect of this outflow. 
As a future work, it remains to construct the model including the effect of the outflow. 
In the phase of the disk evolution, the turbulence arising from the magnet rotational instability (MRI) is considered to become the main source of viscosity. 
In the region where ionization degree is low (called as dead zone), the MRI turbulence is inactive \citep{San00}. 
Thus, in the dead zone, the viscous parameter $\alpha$ is expected to be much smaller than that in the MRI active region. 
Although we used a constant $\alpha$ model in this study, as the future work, it will be important to consider the realistic viscous parameter based on the physical processes with the relations among other physical quantities.

\section{Conclusions}\label{sec:CON}
In this study, the gravitational stability of protoplanetary disks is studied. 
The temperature in a protoplanetary disk was analytically calculated with considering thermal effects such as radiative cooling, viscous heating, and ambient heating source in a molecular cloud. 
It is found that the temperature becomes as low as $T\le$20K in optically thin regime. 
We derived a critical surface density $\Sigma_{\rm c}$, which is necessary for the disk to become unstable, as a function of radius. 
By comparing a given surface density to $\Sigma_{\rm c}$, the possibility and the position of the gravitational instability can be predicted. 
The formation of ices is important for the gravitational instability, and we found that the $\Sigma_{\rm c}$ changes discontinuously at the critical radius $r_{\rm c}\sim$20AU, where ices form.

Two semi-analytic models of protoplanetary disks are used with above critical surface density $\Sigma_{\rm c}$ in order to discuss the possibility and the position of gravitational instability. 
In the model of a steady state accretion disk, which corresponds to the evolved disk, it is found that the disk becomes unstable when the mass accretion rate is greater than the critical mass accretion rate $\dot M_{\rm crit}$ in equation (\ref{MdotC}). 
From this result, it is expected that most of the disks around T-Tauri stars are gravitationally stable and that the disks around protostar are unstable in $r\ge r_{\rm c}$. 
In the model of the disk with the same angular momentum distribution as the former cloud core, which corresponds to the outer region of the young disk, the disk is expected to become unstable in $r\ge r_{\rm c}$ for the case with large rotation parameter ($\beta_0\sim 10^{-3}$). 
From the results of these two models, we conclude that the fragmentation near the central star ($r<r_{\rm c}$) is expected to be rare to occur as long as the parameter range indicated by observations is considered.

\bigskip

We thank Fumio Takahara, Yutaka Fujita, and Hideyuki Tagoshi for fruitful discussion and continuous encouragement. 
We are also grateful to Taishi Nakamoto and Kei Tanaka who provided helpful comments and suggestions.

\appendix
\section{The expression of Temperature, Q value, and Critical Surface Density} \label{app:TCSD}
Here we summarize the value and dependence of the temperature, Toomre Q value, and the critical surface density $\Sigma_{\rm c}$.

\subsection{Equilibrium Temperature}
The equilibrium temperature given in equation (\ref{temp}) is explicitly written below. 
The temerature range for each formula is given in table \ref{tab:BellLin}

\subsubsection{Optically Thick Case}

(a)ices
\begin{equation}
T=12{\rm K} \left(\frac{r}{10\rm AU}\right)^{-3/2}\left(\frac{\Sigma}{10^2\rm g/cm^2}\right)^{2}\left(\frac{M_{\rm s}}{1\rm M_{\odot}}\right)^{1/2}\left(\frac{\alpha}{0.01}\right)^{1}
\end{equation}

(b)sublimation of ices
\begin{equation}
T=1.8\times10^2{\rm K} \left(\frac{r}{1\rm AU}\right)^{-3/20}\left(\frac{\Sigma}{10^2\rm g/cm^2}\right)^{1/5}\left(\frac{M_{\rm s}}{1\rm M_{\odot}}\right)^{1/20}\left(\frac{\alpha}{0.01}\right)^{1/10}
\end{equation}

(c)metal dust
\begin{equation}
T=8.0\times10^2{\rm K} \left(\frac{r}{1\rm AU}\right)^{-3/5}\left(\frac{\Sigma}{10^3\rm g/cm^2}\right)^{4/5}\left(\frac{M_{\rm s}}{1\rm M_{\odot}}\right)^{1/5}\left(\frac{\alpha}{0.01}\right)^{2/5}
\end{equation}

(d)sublimation of metal dust
\begin{equation}
T=1.2\times10^3{\rm K} \left(\frac{r}{1\rm AU}\right)^{-6/55}\left(\frac{\Sigma}{10^4\rm g/cm^2}\right)^{6/55}\left(\frac{M_{\rm s}}{1\rm M_{\odot}}\right)^{2/55}\left(\frac{\alpha}{0.01}\right)^{2/55}
\end{equation}

\subsubsection{Optically Thin Case}
As described in section \ref{sec:TEMP}, the temperature become very low in optically thin case. 
Thus, we consider only the case with ices opacity.

(a)ices
\begin{equation}
T=21{\rm K} \left(\frac{r}{1\rm AU}\right)^{-3/10}\left(\frac{M_{\rm s}}{1\rm M_{\odot}}\right)^{1/10}\left(\frac{\alpha}{0.01}\right)^{1/5}
\end{equation}

\if0

\subsection{光学的深さ}
\ref{sec:T}節で述べたように、氷ダスト以外が主成分のときは常に光学的に厚い。
光学的厚い状態から薄い状態へ変わる点が重要であるため、氷ダストが主成分の場合のみ光学的厚さを求めると
\item[氷ダスト]
\begin{equation}
T=12{\rm K} \left(\frac{r}{10\rm AU}\right)^{-3/2}\left(\frac{\Sigma}{10^2\rm g/cm^2}\right)^{2}\left(\frac{M_{\rm s}}{1\rm M_{\odot}}\right)^{1/2}\left(\frac{\alpha}{0.01}\right)^{1}
\end{equation}
\fi

\subsection{The Q Value}

The Q value defined in equation (\ref{ToomreQ}) is represented below. 

\subsubsection{Isothermal Region}
\begin{equation}
Q=\frac{\Omega c_{\rm s,min}}{\pi G\Sigma}=21\left(\frac{\Sigma}{1\rm g/cm^2}\right)^{-1}\left(\frac{r}{100AU}\right)^{-3/2}
\end{equation}

\subsubsection{Optically Thick Case}
In optically thick case, Q value is represented as 
\begin{equation}
Q=C\left(\frac{27}{256}\kappa_0AB^a\right)^{\frac{1}{6+a-2b}}r^{-\frac{3}{2}\frac{7+2a-2b}{6+a-2b}}\Sigma^{\frac{2b-4}{6+a-2b}},\label{Qthick}
\end{equation}
where A, B, and C is defined in equations (\ref{A}), (\ref{B}), and (\ref{C}), respectively. 
Using the value of table \ref{tab:BellLin}, we can see the dependence of Q value. 
From equation (\ref{Qthick}), it is found that the Q value is independent of surface density for the case with $b=2$, which is realized when ices dominate the opacity. 

\subsubsection{Optically Thin Case}
In optically thin case, Q value is represented as
\begin{equation}
Q=C\left(\frac{27A}{64\kappa_0B^a}\right)^{\frac{1}{6-a+2b}}r^{\frac{3}{2}\frac{2a-2b-7}{6-a+2b}}\Sigma^{\frac{-6-2b}{6-a+2b}},\label{Qthin}
\end{equation}
where A, B, and C is defined in equations (\ref{A}), (\ref{B}), and (\ref{C}), respectively.

\subsection{The Critical Surface Density $\Sigma_{\rm c}$ and the Critical Radius $r_{\rm c}$} \label{App:CSD}

(a)isothermal region
\begin{equation}
\Sigma_{\rm c}=\frac{c_{\rm{s,min}}\Omega}{\pi G}=21\ {\rm g/cm}^2\ \left(\frac{r}{100{\rm AU}}\right)^{-3/2}\left(\frac{T_{\rm{min}}}{10{\rm K}}\right)^{1/2}\left(\frac{M_s}{\rm M_{\odot}}\right)^{1/2} 
\end{equation}
This is available for $r_{\rm c}<r$, where $r_{\rm c}$ is the critical radius represented as equation (\ref{rc}). 

(b)ices

In this regime, unique $\Sigma_{\rm c}$ is not present but $r_{\rm c}$ is present. 
\begin{equation}
r_{\rm c}=\left(\frac{\gamma k_{\rm B} M_{\rm s}^{3/4}}{\pi \overline m G^{1/4}}\sqrt{\frac{27\kappa_0\alpha}{256\sigma}}\right)^{4/9}=24\rm{AU}\left(\frac{\alpha}{0.01}\right)^{2/9}\left(\frac{M_{\rm s}}{\rm M_{{\odot}}}\right)^{1/3}\ 
\end{equation}

(c)sublimation of ices
\begin{equation}
\Sigma_{\rm c}=3.4\times10^3{\rm g/cm^2} \left(\frac{r}{10\rm AU}\right)^{-7/4}\left(\frac{M_{\rm s}}{1\rm M_{\odot}}\right)^{7/12}\left(\frac{\alpha}{0.01}\right)^{1/18}
\end{equation}
This is available in $r_1<r<r_{\rm c}$, where $r_1$ is the radius where the main component of opacity changes between sublimation of ices and ices. 
This $r_1$ is represented as
\begin{equation}
r_1=16{\rm AU}\left(\frac{M_s}{1\rm M_{\odot}}\right)^{1/3}\left(\frac{\alpha}{0.01}\right)^{2/9}
\end{equation}

(d)metal dust
\begin{equation}
\Sigma_{\rm c}=6.2\times10^3{\rm g/cm^2} \left(\frac{r}{10\rm AU}\right)^{-3}\left(\frac{M_{\rm s}}{1\rm M_{\odot}}\right)^{1}\left(\frac{\alpha}{0.01}\right)^{1/3}
\end{equation}
This is available in $r_2<r<r_1$, where $r_2$ is the radius where the main component of opacity changes between metal dust and sublimation of ices. 
This $r_2$ is represented as
\begin{equation}
r_2=10{\rm AU}\left(\frac{M_s}{1\rm M_{\odot}}\right)^{48/141}\left(\frac{\alpha}{0.01}\right)^{98/423}
\end{equation}

(e)sublimation of metal dust
\begin{equation}
\Sigma_{\rm c}=6.3\times10^3{\rm g/cm^2} \left(\frac{r}{10\rm AU}\right)^{-171/104}\left(\frac{M_{\rm s}}{1\rm M_{\odot}}\right)^{7/13}\left(\frac{\alpha}{0.01}\right)^{1/52}
\end{equation}
This is available in $r_{\rm edge}<r<r_2$, where $r_{\rm edge}$ is the inner most radius for the gravitational instability represented as
\begin{equation}
r_{\rm{edge}}=1.2{\rm AU}\left(\frac{M_{\rm s}}{\rm M_{\odot}}\right)^{1/3}\left(\frac{\alpha}{0.01}\right)^{54/353} 
\end{equation}
We explain this $r_{\rm edge}$ in Apendix \ref{app:redge}

\section{The inner most raius for the gravitational instability}\label{app:redge}
In the regime where molecules mainly contribute to the opacity, which is realized at $T\gtrsim1600$ K, the temperature dependence of cooling rate becomes weak as $\Lambda\propto T^{4/3}$. 
Then, the temperature becomes too high with large surface density, and the stabilizing effect of thermal pressure becomes larger than the destabilizing effect of surface density increasing. 
It means that increasing surface density stabilizes the disk. 
Thus, the line of $\Sigma_{\rm c}$ become multi-valued function of radius. 
In figure \ref{fig:Qcntr}, we see this feature for $\Sigma\sim2\times10^5\rm g/cm^2$ and $r\sim1$AU. 
The upper line of $\Sigma_{\rm c}$ slopes upwards when going from left to right. 
This is due to the lower heating rate at larger $r$. 
Because the $ \Sigma_{\rm c} $ line slopes downwards where the main component of opacity is sublimation of dust grains, the unstable region has the inner edge at the point where the main component of opacity changes between molcules and sublimation of dust grains. 
In figure \ref{fig:Qcntr}, we see that this inner edge locates at $r \simeq 1.2 {\rm AU} $ and $ \Sigma \simeq 2 \times 10^5 {\rm g/cm^2}$. 
We name this radius $r_{\rm edge}$. 
For $r<r_{\rm edge}$, the disk cannot become gravitationally unstable no matter how large the surface density becomes. 
By using the equation (\ref{temp}), (\ref{SigmaCrit}), and the condition for opacity-change from dust sublimation to molecules, $r_{\rm edge}$ is represented as
\begin{equation}
r_{\rm edge}=1.2{\rm AU}\left(\frac{M_{\rm s}}{\rm M_{\odot}}\right)^{1/3}\left(\frac{\alpha}{0.01}\right)^{54/353}\ . \label{redge}
\end{equation}
This $r_{\rm edge}$ weekly depends on $\alpha$ and $M_{\rm s}$ just like a case of $r_{\rm c}$. 
Thus, even if $\alpha$ or $M_{\rm s}$ is much different from typical value, the value of $r_{\rm edge}$ varies only severalfold.

\if0

\section{The AMC model with the sphere of uniform density} \label{app:UD}
We also perform the calculation for the case that the former core is the sphere of uniform density. 
In this case, the core is assumed to be uniform density and rigidly rotating. 
The sphere of uniform density is characterized by two parameter. 
One is the rotation parameter $\beta_0\equiv E_{\rm rot}/|E_{\rm grav}|$ and the other is the thermal parameter $\alpha_0\equiv E_{\rm th}/|E_{\rm grav}|$, where $E_{\rm th}$ is the thermal enery. 
In the case with the sphere of uniform density, these parameter is represented as
\begin{equation}
\alpha_0=\frac{5 R_{\rm E}c_{\rm s,0}^2}{2GM_{\rm c}}
\end{equation}
and
\begin{equation}
\beta_0=\frac{\Omega_0^2R_{\rm E}^3}{3GM_{\rm c}}.\label{udbeta0}
\end{equation}
We set the mass of core $M_{\rm c}=1\rm M_{\odot}$, the sound speed $c_{\rm s}=190\rm m/s$. 
Then, the cloud radius $R_{\rm E}$ is determined for a given $\alpha_0$. 
Here we fix $\alpha =0.86$, which is the same value for the case of the Bonnor-Ebert sphere. 
The angular velosity $\Omega_0$ is determined from equation (\ref{udbeta0}). 
Then, we can calculate the surface density of the disk for a given the ratio of disk to star mass $M_{\rm d}/M_{\rm s}$.

\section{Approximation of ...}

\section*{Complete data}
\fi

%%%
% See the manual for the detail.
%%%

\end{document}